\documentclass[12pt,preprint]{aastex}
\usepackage{epsfig}

\newcommand{\eqb}{\begin{eqnarray}}
\newcommand{\eqe}{\end{eqnarray}}

\newcommand{\rhoext}{\rho_{\rm CSM}}

\newcommand{\qprinj}{Q_{p}^{\rm inj}}
\newcommand{\qelinj}{Q_{e}^{\rm inj}}

\def\gsim{\mathrel{\raise.5ex\hbox{$>$}\mkern-14mu
             \lower0.6ex\hbox{$\sim$}}}

\def\lsim{\mathrel{\raise.3ex\hbox{$<$}\mkern-14mu
             \lower0.6ex\hbox{$\sim$}}}

\slugcomment{submitted to ApJLetters, 2008}

\shorttitle{GRB}
\shortauthors{Mastichiadis \& Kazanas}

\begin{document}


\title{The Supercritical Pile GRB Model: The Prompt to Afterglow
Evolution}


\author{A. Mastichiadis}
\affil{Department of Physics, University of Athens, Panepistimiopolis,
  GR 15783, Zografos, Greece}
\author{D. Kazanas}
\affil{Astrophysics Sciences Division, NASA/GSFC, Code 663,
Greenbelt, MD 20771}



\begin{abstract}

The ``Supercritical Pile" is a very economical GRB model  that
provides for the efficient conversion of the energy stored in the
protons of a Relativistic Blast Wave (RBW) into radiation and at the
same time produces - in the prompt GRB phase, even in the absence of
any particle acceleration - a spectral peak at energy $\sim 1$ MeV.
We extend this model to include the evolution of the RBW Lorentz
factor $\Gamma$ and thus follow its spectral and temporal features
into the early GRB afterglow stage. One of the novel features of the
present treatment is the inclusion of the feedback of the GRB
produced radiation  on the evolution of $\Gamma$ with radius. This
feedback and the presence of kinematic and dynamic thresholds in the
model are sources of potentially very rich time evolution which we
have began to explore. In particular, one can this way obtain
afterglow light curves with steep decays followed by the more
conventional flatter afterglow slopes, while at the same time
preserving the desirable features of the model, i.e. the well
defined relativistic electron source and  radiative processes that
produce the proper peak in the $\nu F_{\nu}$ spectra. In this note
we present the results of a specific set of parameters of this model
with emphasis on the multiwavelength prompt emission and transition
to the early afterglow.

\end{abstract}




\pagebreak

\section{Introduction}
The cosmological origin of GRB has by now been firmly established
following the discovery of their afterglows and the determination of
their redshifts \citep{costa, vanP} and the launch of {\sl Swift}
which increased the number of observed afterglows and redshift
determinations. These developments left little doubt that GRB
emission is intimately associated with Relativistic Blast Waves
(RBW), as proposed by \citet{rees} and at the same time shifted the
focus of the study from the prompt GRB emission to its afterglow
\citep{zhang, piran}.

The early, sparsely sampled GRB afterglow light curves, were fit
well with simple power law functions, appropriate to emission from
either spherical \citep{sarpinar} or jet-like \citep{sarpihal} RBW.
However, the launch of {\sl Swift} with its prompt, continuous,
broad frequency coverage has provided new unexpected (and
unexplained) details of the afterglow light curves. Chief amongst
them are: (a) { An early afterglow steep decrease of the flux
{\rm($\propto t^{-3}$ to $t^{-6}$)} followed often by a period of
constant flux} (before its eventual power law decline) in many
bursts. (b) { Large flares in the X-ray light curves $\sim
10^3-10^5$ sec after the beginning of the event}
 \citep[see][for more details]{obrien}. These were compounded to
the already open problems of the prompt emission, namely: (c) { The
GRB ``inner engine"}. (d) {The non-dissipative transport of the GRB
energy to the emission region and, most importantly, its efficient
dissipation there}. (e) {The physics behind the characteristic
energy of peak GRB emission, $E_{\rm p}$ and its narrow distribution
within the class of the classic GRB} \citep[]{malozzi, preece}. (f)
{The physics that relate GRB to XRR (X-Ray Rich bursts)  and XRF
(X-Ray Flashes), transients of lower flux and lower $E_{\rm p}$},
recorded by broad band missions such as {\sl BeppoSAX}, {\sl HETE}
and {\sl Swift} \citep[e.g][]{yonet04}.

Of the above problems, (a) has received no apparent resolution while
(b) is loosely attributed to continued activity at the ``inner
engine"; while not implausible, this demands activity over time
scales almost $10^7$ times longer than the characteristic time
associated with the ``inner engine" dynamics ($\simeq 10^{-3}$ sec),
as the latter is thought related to stellar collapse. (d) is
considered to be effected either through protons \citep[e.g][]{rees}
or magnetic fields \citep{vlah}, however, the necessary and
efficient dissipation ``is one of the least studied aspects of GRB"
\citep{piran}; this issue is generally approached by {
parameterizing} the energy density in relativistic electrons to be a
given fraction (typically $\sim 50\%$) of that of protons. Issue (e)
is generally open, given the absence of an underlying reason for
such a characteristic energy. Monte Carlo simulations of a large
number of models \citep{zhang2} failed to reproduce the narrow width
of the observed distribution because of the large number of
parameters involved and/or because of the lack of strong dependence
of $E_{\rm p}$ on any single parameter. Finally, there are a number
of proposals concerning (f) \citep{ioka,dcb}, which appear plausible
but without  a single one of them universally agreed upon.

The ``Supercritical Pile" Model (SPM) \citep[henceforth
KGM02 and MK06]{kazan02,maskaz}, adapted from AGN \citep{kazmast}, has been
introduced to provide a resolution to (d). The compelling arguments
in favor of the SPM are: (1) {\em Its economic} (non-thermal
particles not necessary), {\em efficient conversion of the RBW
relativistic proton energy into photons} through a radiative
instability akin to that of a supercritical nuclear pile. (2) {\em
Its spectra which exhibit a characteristic value for $E_{\rm p}
\simeq 1$} MeV (in the lab frame) irrespective of the RBW Lorentz
factor $\Gamma$, in agreement with observation \citep{malozzi}, {\em
produced as ``unintended consequence" of the dissipation
process}. Crucial in addressing these issues has been the presence
of an upstream medium which scatters the RBW photons (a ``mirror")
and allows them to be re-intercepted by the RBW, while boosted in
energy by $\simeq \Gamma^2$.

{ In MK06 we have explored numerically the SPM assuming a constant
Lorentz factor $\Gamma$ for the RBW, confirmed the efficiency of
proton energy conversion into to radiation and the presence of a
well defined value for $E_{\rm p}$, reflecting the kinematic
threshold of the reaction $p \gamma \rightarrow p \, e^+e^-$. The
present treatment is far more realistic: (a) It computes the
evolution of the RBW Lorentz factor $\Gamma$ through a medium of
density $n(r) \propto R^{-2}$, thought to represent the wind of a WR
star, including also the effects of the radiative drag of the
bulk-Comptonized photons. (b) Replaces the upstream ``mirror"
required by the model by scattering the RBW photons in this medium.
The combination of these effects can result in a rich GRB time
evolution,  but we presently restrict ourselves to a specific
example of a GRB light curve,} deferring the broader exploration of
other models to a future publication. { Despite this limited scope,
we can reproduce some} of the salient features of the GRB in the
afterglow evolution, such as their steep decrease in flux following
the termination of their prompt phase, an effect traceable in this
specific case to the kinematic threshold of the model.

In \S 2 we provide the general framework of our model with emphasis
on its novel aspects compared to previous treatments. In \S 3 we
present the results of our calculations and finally in \S 4 the
results are summarized and conclusions are drawn.

\section{The Coupled Radiative -- Dynamical Evolution}


We consider a Relativistic Blast Wave (RBW) of speed
$\upsilon_{0}=\beta_\Gamma c$ and Lorentz factor $\Gamma$. Its
radius $R(t)$ is measured from the center of the original explosion and it is
sweeping up the CircumStellar Medium (CSM) of density $\rhoext$. The
evolution of $\Gamma$ as a function of radius is given by the
combination of the conservation laws of mass
\begin{equation}
{dM\over{dR}}=4\pi R^2\Gamma\rhoext - {1\over{c^3\Gamma}}\dot E
\label{mass}
\end{equation}
and energy-momentum
\begin{equation}
{d\Gamma\over{dR}}=-{4\pi R^2\rhoext\Gamma^2\over
M}-{F_{\rm rad}\over{Mc^2}}.
\label{energy}
\end{equation}
\citep{cd99}. Here $\dot E$ is the radiation emission rate  as
measured in the comoving frame and $F_{\rm rad}$ is the radiation
drag force exerted on the RBW by any radiation field exterior to the
flow. Given that the RBW velocity $v_0$ is very close to the speed
of light $c$, the entire radiative history of the RBW lies just
ahead of it at a distance $D \sim R/\Gamma^2$; therefore, isotropization
of this radiation by scattering in the ambient medium (the action of the
``mirror") will lead to its re-interception by the RBW to thus contribute
to $F_{\rm rad}$. This is given by the expression
\begin{equation}
F_{\rm rad}={{64\pi}\over{9c}}\tau_b n_e^{CSM}\sigma_T
R\Gamma^4\dot{E}
\label{rad}
\end{equation}
\noindent where $\tau_b$ is the RBW Thomson depth, $n_e^{CSM}= \rhoext/m_H$
the CSM electron density and $\sigma_T$ the Thomson cross section. In the
above expression, two powers of $\Gamma$ are due to the increase of
the photon energy density upon its scattering on the ``mirror" while
the other two to the usual radiative loss rate (an analogous term
due to the $p \gamma \rightarrow pe^+e^-$ reaction was found to increase
$F_{\rm rad}$ by 20\% for the specific parameter values
discussed herein but it maybe more important for different values).
%
The calculation of the $\dot E$ and $F_{rad}$ terms is done by
implementing the numerical code used in MK06 to compute the
radiation of the SPM. This is done be solving the simultaneous
equations
\begin{equation}
{{\partial n_i}\over{\partial t}} + L_i +Q_i=0.
\label{continuity}
\end{equation}
The unknown functions $n_i$ are the differential number densities of
protons, electrons and photons while the index $i$ can be any one of
the subscripts `p', `e' or `$\gamma$' referring to each species. The
operators $L_i$  denote losses or escape of each species from the
system while $Q_i$ denote injection and source terms of each species
by each of a number of processes which are described in detail in
MK06. The above equations are solved in the fluid frame in a
spherical volume of radius $R_b=R/\Gamma$. This can be justified by
the fact that due to relativistic beaming an observer receives the
radiation coming mainly from a small section of the RBW of lateral
width $R/\Gamma$ and longitudinal width $R/\Gamma^2$ in the lab
but $R/\Gamma$ on the comoving frame.

The present treatment differs from that of MK06 in two important aspects:

(1) Hot protons accumulate continuously on the RBW as it sweeps the CSM.
This then sets the source terms of the protons ($\qprinj$) and electrons
($\qelinj$) (with units particles/energy/volume/time) to
\begin{equation}
\qprinj={{\rhoext c}\over{m^2_pc^2R}}(\Gamma^2-\Gamma)\delta(\gamma_p-\Gamma)
\label{Qp}
\end{equation}
and
\begin{equation}
\qelinj={{\rhoext c}\over{m_pm_ec^2R}}(\Gamma^2-\Gamma)\delta(\gamma_e-\Gamma)
\label{Qe}
\end{equation}
i.e. we assume that at each radius $R$ the RBW picks up an equal
amount of electrons and protons from the circumstellar medium  which
have, upon injection,  energies $E_p=\Gamma m_pc^2$ and $E_e=\Gamma
m_ec^2$ respectively. Consequently, the proton {\sl energy}
injection rate is given by \citep{bmckee}
\begin{equation}
\left({{dE}\over {dt}}\right)_{\rm inj}=4\pi R^2\rhoext(\Gamma^2-\Gamma)c^3
\label{Qtot}
\end{equation}
while a fraction $m_e/m_p$ of the above goes to electrons.

(2) The scattering of the RBW photons takes place on the CSM ahead
of the advancing RBW (rather than an {\em ad hoc} mirror) and, as
such, its photon scattering column is uniquely determined by the
initial conditions and, like all other parameters is a function of
time (or equivalently position).

Eqns. (\ref{mass}), (\ref{energy}), (\ref{continuity}), along with
Eqns. (\ref{rad}), (\ref{Qp}) and (\ref{Qe}) form a set
which can be solved to yield simultaneously the evolution of
the RBW dynamics and luminosity.
This approach is self-consistent in that the `hot' mass injected
through equations (\ref{Qp}) and (\ref{Qe}) shows up at RHS of Eqn.
(\ref{mass}), while the radiated luminosity $\dot E$ feedbacks onto
the energy-momentum equation through the definition of the radiative
force, $F_{\rm rad}$, of Eqn. (\ref{rad}).
The free parameters of this system are (i) the total energy of the
explosion $E_{\rm tot}$ (ii) the CSM density profile $n(r)$ (iii)
the magnetic field as a function of radius $B(r)$.
To avoid computation of the evolution during the RBW acceleration phase
when it likely produces little radiation, we have chosen to
begin our calculations (and the accumulation of matter by the RBW)
at a radius $R_0$ at which is has already achieved its asymptotic
Lorentz factor $\Gamma_0 = \Gamma (R_0)$.

 \begin{figure}[t]
 \vspace*{-5.0mm} 
 \includegraphics[width=7.0cm, angle=-90]{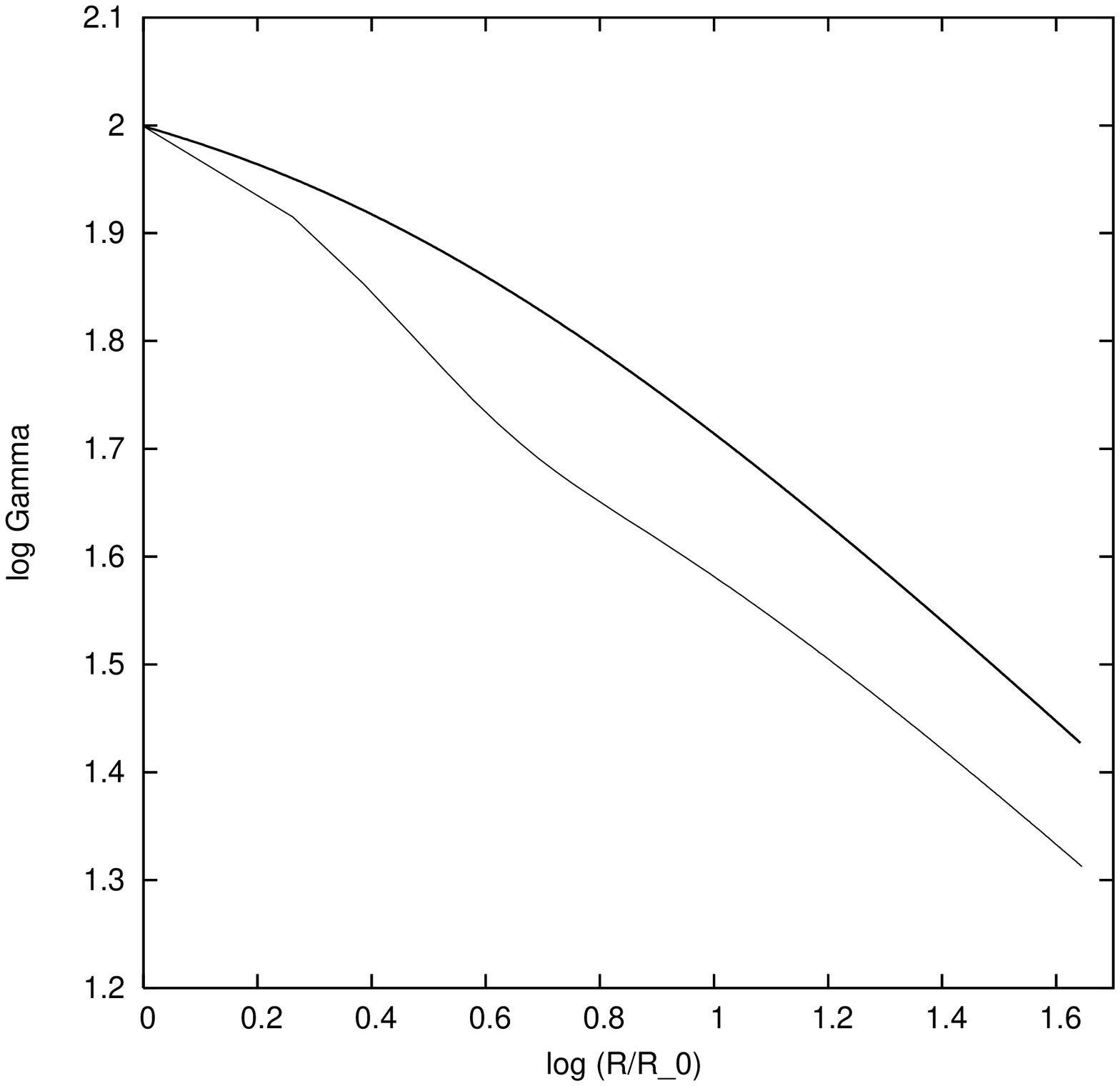}
\hskip -12pt
 \includegraphics[width=7.0cm, angle=-90]{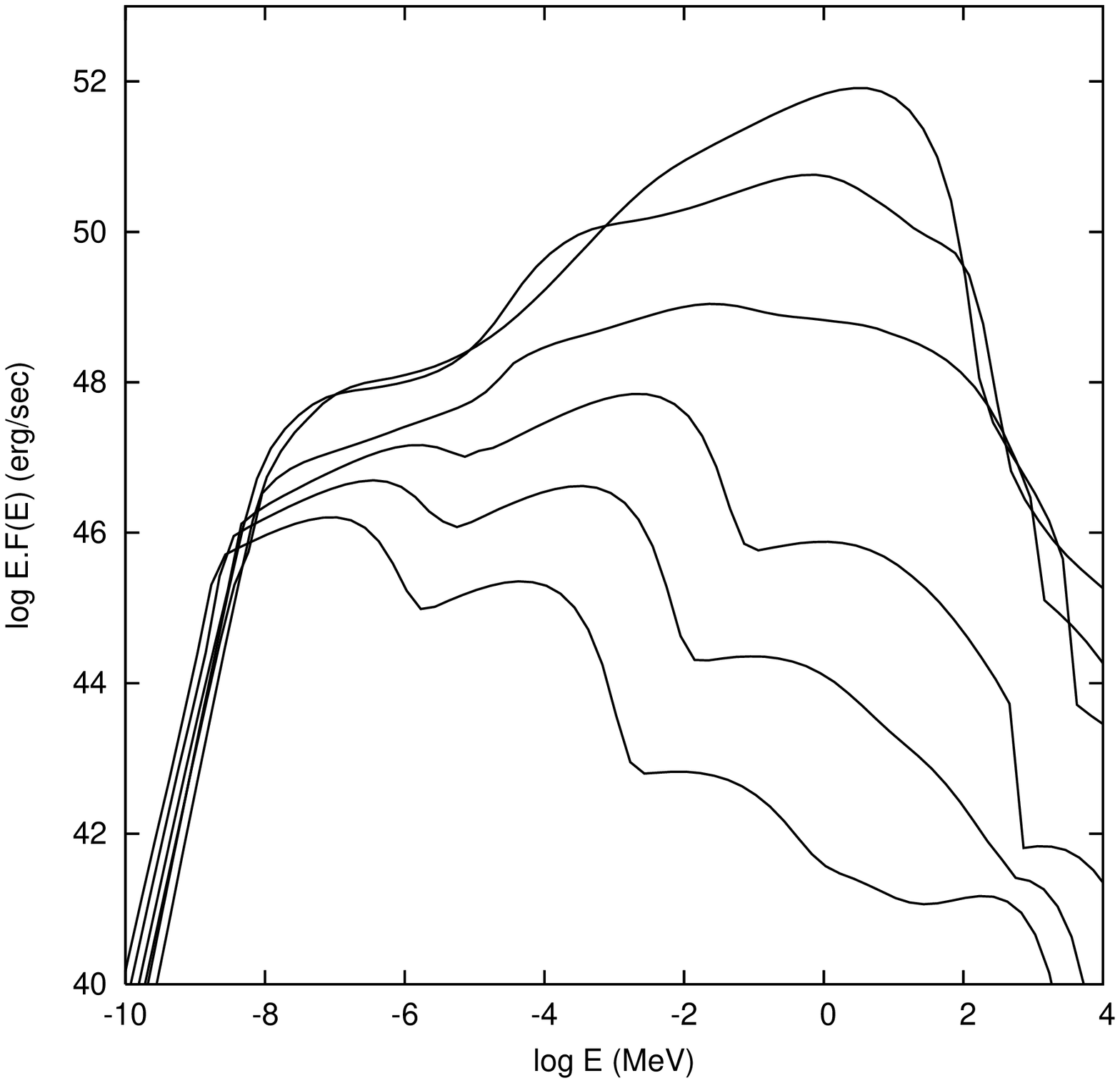}
%
 \caption{\footnotesize (a) The Lorentz factor  for  a RBW propagating in
a wind environment with parameters given in the text. 
The thick lines show the evolution with radius without radiative
drag, while the thin one with the drag included. (b) The
multiwavelength spectrum of the burst at times 1, 3, 10, 30, 100 and
300 sec (top to bottom). The peak of the bulk Comptonized component
is originally close to 1 MeV, however, as the evolution proceeds it
moves to lower energies.} \label{fig1}
 \end{figure}

 \begin{figure}[t]
 \vspace*{-5.0mm} 
 \includegraphics[width=7.0cm, angle=-90]{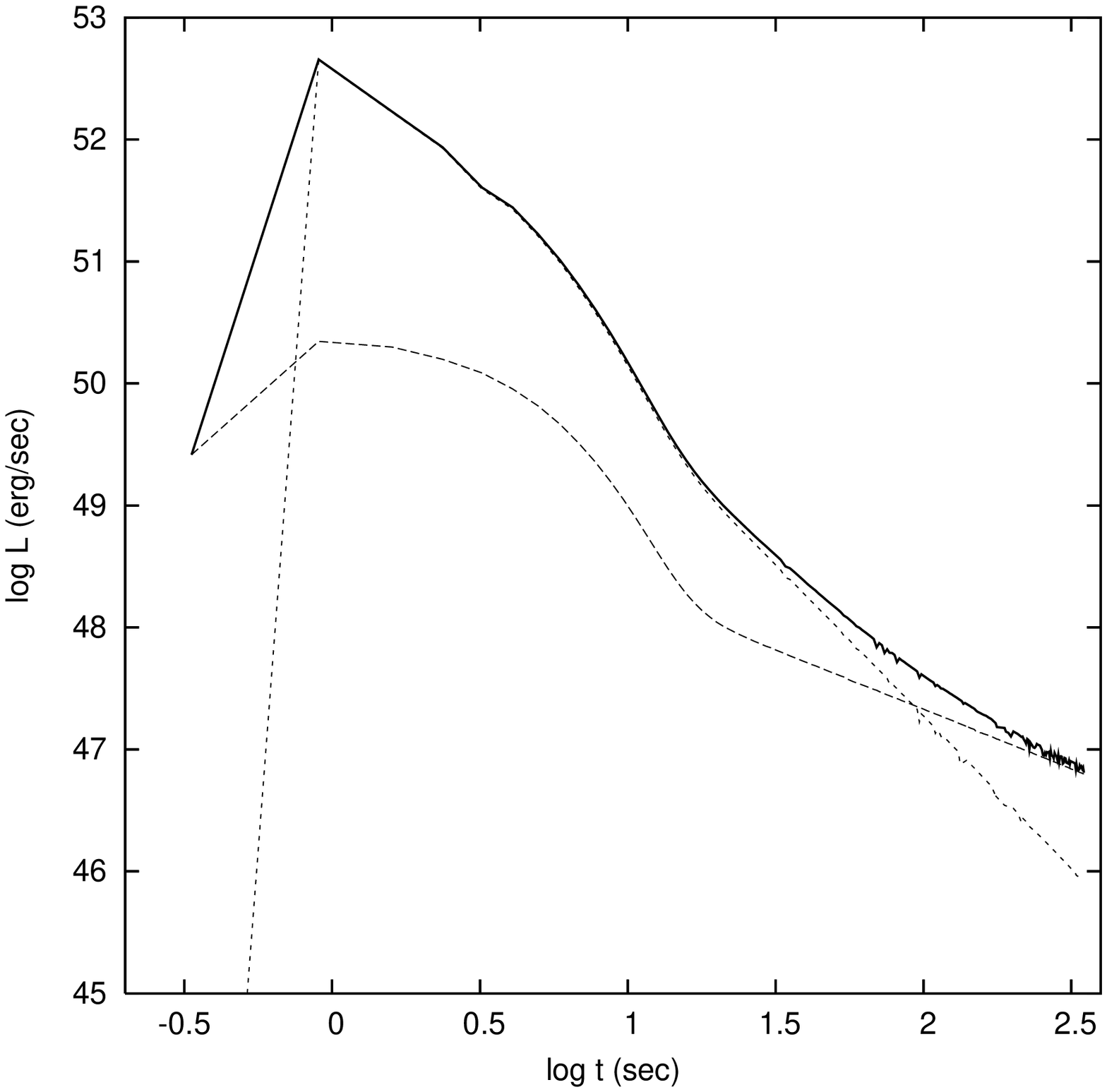}
\hskip -12pt
 \includegraphics[width=7.0cm, angle=-90]{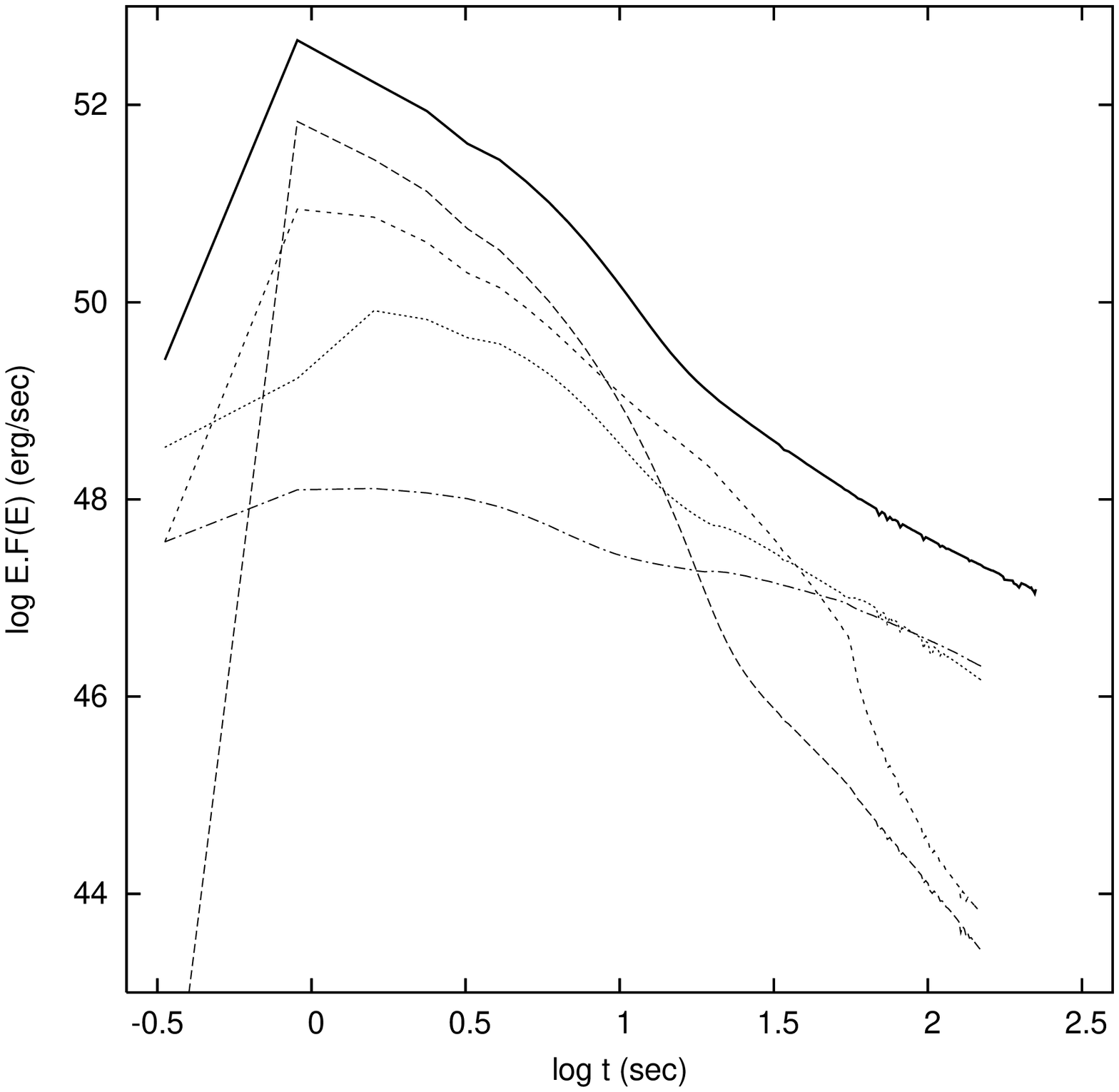}
%
 \caption{\footnotesize (a)
The bolometric burst lightcurve. The dashed line corresponds to
the internal RBW luminosity while the dotted line is the
bulk Comptonized component. The thick line is the sum of the two.
(b)
The corresponding burst luminosity for various energy bands
as a function of time. Long dashed line is at energy of 1 MeV,
short dashed is at 10 keV, dotted at 100 eV and dot-dashed at 1 eV.
The thick full line is the bolometric lightcurve.
Parameters are as given in the text.}
\label{fig2}
 \end{figure}

%
As proposed in KGM02 and shown explicitly in MK06, the relativistic
protons accumulated in the RBW can become supercritical to the
network of $p \gamma \rightarrow p \, e^+ e^-, ~e B \rightarrow
\gamma$ once kinematic and dynamic thresholds are simultaneously
fulfilled. The kinematic threshold simply reflects the kinematic
threshold of the $p \gamma \rightarrow p \, e^+ e^-$ reaction and
reads
\begin{equation}
b \, \Gamma^5\gsim 1 ~~{\rm or~ }~~ \Gamma \gsim 214
~(n_0)^{-1/12}  ~~{\rm for} ~~ B=B_{eq} \simeq (8\pi m_p c^2 n_0)^{1/2} 
\Gamma
\label{thres1}
\end{equation}
with the latter expression for B-field in equipartition, where
$b=B/B_{crit}$ and $B_{crit}=m^2c^3/e\hbar$ is the critical magnetic
field. The dynamic threshold provides the critical column density
for the accumulated relativistic protons to become supercritical (in
a fashion analogous to a nuclear pile) and, if fulfilled, a large
fraction of the energy stored in relativistic protons is converted
into e$^+$e$^-$-pairs within a few light travel times across the
width of the shock.

At the earliest stages of the RBW evolution the accumulated
relativistic proton column is small and little emission is possible,
only that of the swept-up electrons, which is smaller than the
energy flux through the shock by a factor $m_e/m_p$ and may very
well represent the oft quoted GRB precursor emission. The eventual
evolution of the RBW depends on whether its asymptotic Lorentz
factor $\Gamma_0$ and $B-$field satisfy the kinematic threshold
(Eqn.  \ref{thres1}). If not, and in the absence of an accelerated
population of particles, only the energy flux in electrons is
converted to radiation and the GRB is a ``dud", as the combination
$b \, \Gamma^5$ is only expected to decrease with radius (however an
explosive release is still possible if the proton distribution
includes an accelerated power law component that extends to $E \gg
\Gamma m_pc^2$; as hinted in \cite{kmg06}, these events may be
related to the XRRs and XRFs).

Far more interesting is the case where the kinematical criterion is
satisfied initially. Then whether the flow becomes radiatively
unstable depends on the
column of hot protons accumulated on the RBW. When this exceeds the
critical value, the energy contained in relativistic protons is
explosively released, the value of $F_{\rm rad}$ increases
dramatically and the Lorentz factor $\Gamma$ of the RBW can decrease
over a distance $D \ll R(t)$, provided that the second term in Eq.
(\ref{energy}) is dominant. This drop in $\Gamma$ is important not
only for decreasing the emitted flux but, more significantly, for
potentially pushing $b \, \Gamma^5$ below its kinematic threshold
value $\simeq 1$, (as is the case shown in Fig. \ref{fig1}a), a fact
that according to the SPM marks the end of the prompt GRB emission
phase, i.e. the conversion of proton energy into radiation.
Following this event, radiation is emitted only from cooling the
electrons already present within the RBW and those being swept-up by
it. The observed flux suffers a precipitous decrease with further
evolution that depends on the value of $R$ relative to the
deceleration radius $R_{\rm dec}$ corresponding to the resulting
value of $\Gamma$; if $R < R_{\rm dec}$, the flux remains at a
roughly constant level until $\Gamma$ resumes its decline, at $R >
R_{\rm dec}$, through accumulation of mass on the expanding RBW; if
$R > R_{\rm dec}$, then $\Gamma$ will continue its decline at the
much slower conventional level of afterglow theory.

The detailed, long term evolution of the GRB flux depends on $E_{\rm
tot}$, $n(r)$ and $B(r)$ that determine the values of $r$ and $\Gamma$
at which the RBW becomes supercritical -- it is
conceivable that for certain parameter combinations supercriticality
can be reached at more than one radius, with the released energy
being proportional to the time between the corresponding bursts see
e.g. \citet{RRM01}. In Figure \ref{fig1} and \ref{fig2} we present
the evolution of a RBW with $n = n_0(R_0/R)^2$ and $B=B_0(R_0/R)$.
The parameters are $R_0 = 10^{14}$ cm, $n_0 = 8.10^8 ~{\rm cm}^{-3},
~\Gamma_0 = 100, ~B_0 = 4.4~10^4$ G and total isotropic energy
$E_{\rm tot} =10^{54}$ erg.

Figure \ref{fig1}a depicts the evolution of $\Gamma$ as a function of
radius in this medium with (thin line) and without (thick line) the radiative
feedback. The  drop in $\Gamma$ corresponds to the explosive energy
release in the protons and the slow down of the RBW due to the
radiation drag. As deduced from this figure, $R_0 \simeq R_{\rm
dec}$, since for $R > R_0$, $\Gamma \propto R^{-1/2}$, as expected
for adiabatic propagation in a wind density profile (thick line).
After the decrease in $\Gamma$ due to the radiative feedback and
after the non-adiabatic effects have died out, the evolution of
$\Gamma$ follows a similar track of lower normalization.

Figure \ref{fig1}b shows the multiwavelength spectra at various
instances as perceived by the observer. As it was shown in MK06 the
spectrum consists of two components, one that is due to the primary
particle emission by particles on the RBW and one due to the bulk
Comptonization of the upstream-reflected primary radiation by the
cold pairs of the RBW. This latter component peaks early on at
~1MeV, but as the burst evolves moves to lower energies since both
$\Gamma$ and $B$ drop outward.

Figure \ref{fig2}a shows the corresponding apparent isotropic
bolometric luminosity  as a function of time. This consists of the
internally produced luminosity ({dashed}) and that due to bulk
Comptonization of the mirror-scattered radiation by the RBW
({dotted}) with the thick line representing their sum. As it can
also be seen from Fig \ref{fig1}b, most of the luminosity, is by far
contained in the bulk-Comptonized component (at $E \sim 1$ MeV) and
exhibits the steepest decrease due to the decrease in $\Gamma$ and
the arrest of additional pair injection from the protons. At longer
time scales, the only injection available is that of the ambient
electrons and the emission exhibits the  $\propto t^{-1}$ behavior
of ``standard" afterglows.

Finally Fig. \ref{fig2}b depicts the luminosity at various energy
bands as a function of time -- here we make no distinction between
the direct and the bulk Comptonized component, but instead we
exhibit their sum. As a rule higher frequencies dominate
more at the early stages of the burst but drop faster due
to a combination of faster cooling and the decrease in $\Gamma$.
This is consistent with observations: the BAT flux (that receives
its major contribution from the bulk Comptonized component)
decreases much faster than the flux in the other bands and its level
defines, in effect, the prompt GRB phase (see also next section).

\section{Summary, Discussion}

We have presented above a first attempt at an integrated version of
the SPM, complete with the coupled RBW dynamics, radiation
production
and accumulation of hot protons  on the RBW from the swept-up matter.
The latter process is fundamental as the increase of the hot proton
column to supercritical values is necessary for the explosive energy
release seen in GRB. Another important feature is the coupling of
the radiation to the dynamics of the RBW, the cause of the abrupt
decrease in $\Gamma$ seen in Fig. \ref{fig1}a. { Because this can
reduce $\Gamma$ below the SPM kinematic threshold, it can
severely reduce the observed flux, especially its
bulk-Comptonized spectral component that peaks at $E_p \simeq 1$ MeV
and constitutes the main GRB channel}. The existence of the
kinematic threshold value for $\Gamma$ and its intimate association
to the radiation emission near $E_p~ (\sim 1$ MeV, the defining GRB
property), affords for the SPM an operational definition of the GRB
prompt phase, a feature unique amongst GRB models: as such, {\em the
prompt GRB phase is the stage in its evolution during which the
kinematic threshold condition of Eq. (\ref{thres1}) is fulfilled},
accompanied by severe reduction in the GRB flux following this stage, as
observed.

The time evolution of the flux in Fig. \ref{fig2} bears great
resemblance to that of many {\sl Swift-XRT} GRB, that exibit a very
steep declining profile followed by a less steep or flat section in
their light curves \citep{obrien}, related, as discussed above, to
the relation between $R_0$ and $R_{\rm dec}$.  We believe that the
straightforward way that the SPM addresses these
vexing for the standard model questions attests to its
relevance to the GRB underlying physics and phenomenology. It should
be noted at this point that the efficiency of conversion of kinetic
energy to radiation depends on the value of the ambient density
$n_0$. This dependence comes through the dynamic threshold of the
SPM, as $n_0$  determines also the value of the upstream albedo
(i.e. of the ``mirror", whose assumption is now obviated), to which
the dynamic threshold is proportional. We plan to explore the
effects of this parameter on the GRB properties in a future
publication.

The duration of the burst shown in Fig. \ref{fig2} is of order of a
few seconds. As such it would be classified as a short burst,
despite the fact that the RBW is assumed to propagate in a medium
with properties akin to the wind of a WR star. We therefore have
presented an explicit model that produces a short burst from an
object of a young stellar population. While it was originally
proposed and supported by the earlier observations that short bursts
are associated with old stellar populations (implying neutron star
collisions as their source of energy), it was shown \citep{berger08}
that $\lsim 1/3$ of them are in fact associated with stellar
populations similar to those of the long GRBs.

The outlook from this first time-dependent treatment of the SPM
replete with the CSM distribution and radiation emission and
feedback is that this model can potentially produce a great variety
of GRB light curves (in agreement with GRB phenomenology) which it
can relate to {\em global} parameters of the system. We plan to
explore thoroughly these parameters in future publications.

\begin{acknowledgments}

This research was funded in part by a Grant from the Special Funds
for Research (ELKE) of the University of Athens.
DK would like to acknowledge support
by an INTEGRAL GO grant.
\end{acknowledgments}

{}

\end{document}